\documentclass{article}
\usepackage{amsmath}
\usepackage{amssymb}
\usepackage[margin=1in]{geometry}
\usepackage{comment}
\usepackage[font=footnotesize, width=0.9\textwidth, labelfont=bf]{caption}
\DeclareMathOperator*{\argmax}{argmax}

\usepackage{graphicx}
\usepackage[numbers]{natbib}

\title{Calibration of Human Driving Behavior and Preference\\ Using Naturalistic Traffic Data}
\author{Qi Dai, Di Shen, Jinhong Wang, Suzhou Huang,\, Dimitar Filev\thanks{Corresponding Author.} \\ 
\{qdai2,\,sdi,\,jwang356,\,shuang10,\,dfilev\}@ford.com \\ 
Ford Motor Company, Dearborn, MI 48126}
\begin{document}

\maketitle

\begin{abstract}
	Understanding human driving behaviors quantitatively is critical even in the era when connected and autonomous vehicles and smart infrastructure are becoming ever more prevalent. This is particularly so as that mixed traffic settings, where autonomous vehicles and human driven vehicles co-exist, are expected to persist for quite some time. Towards this end it is necessary that we have a comprehensive modeling framework for decision-making within which human driving preferences can be inferred statistically from observed driving behaviors in realistic and naturalistic traffic settings. Leveraging a recently proposed computational framework for smart vehicles in a smart world using multi-agent based simulation and optimization, we first recapitulate how the forward problem of driving decision-making is modeled as a state space model. We then show how the model can be inverted to estimate driver preferences from naturalistic traffic data using the standard Kalman filter technique. We explicitly illustrate our approach using the vehicle trajectory data from Sugiyama experiment that was originally meant to demonstrate how stop-and-go shockwave can arise spontaneously without bottlenecks. Not only the estimated state filter can fit the observed data well for each individual vehicle, the inferred utility functions can also re-produce quantitatively similar pattern of the observed collective behaviors. One distinct advantage of our approach is the drastically reduced computational burden. This is possible because our forward model treats driving decision process, which is intrinsically dynamic with multi-agent interactions, as a sequence of independent static optimization problems contingent on the state with a finite look ahead anticipation. Consequently we can practically sidestep solving an interacting dynamic inversion problem that would have been much more computationally demanding. We further demonstrate how the calibrated model can be used to gain insights on human driving behaviors via counterfactual simulations. A number of variations of the modeling setting show that our approach is robust and intuitively generalizable.
\end{abstract}
 
\newpage
\section{\bf Introduction}\label{Introduction}
Modeling human driving behaviors in a variety of traffic scenarios is both challenging and important. Even at the advent of new technologies, such as autonomy and connectivity, this endeavor is as relevant as ever. There are multiple reasons for this, a few of which are enumerated here. First, the autonomous vehicle technologies are yet to be sufficiently mature to completely replace human driven vehicles. Therefore, mixed traffic, where autonomous vehicles and human driven ones co-exist, will persist for a long time. Second, even when autonomous vehicles become dominating, the messengers will still be human whose riding experience needs to be respected. This in turn implies that autonomous vehicles have to be able to mimic closely human driving behaviors, at least for a while initially. Third, due to the diversity of individual human driver, offline modeling heterogeneity of human driving in general and online adapting to personal driving style in particular are highly desirable. Fourth, it will be extremely helpful for autonomous vehicle technology development if we were equipped with good human driving behavioral models under a variety of traffic settings. Lastly, most of the autonomous technologies and literature are vehicle-centric, in the sense that they typically aim at successfully steering a single vehicle while treat other road users at the scene almost passively. A consequence of this narrow strategy is that collective behaviors, such as traffic congestions and safety, are often not explicitly tackled. It is this type of observations that motivate our work. We attempt to construct a modeling framework so that all these issues can be addressed simultaneously.

Roughly, there are two broad classes of approaches to modeling human driving behaviors: one is policy based, and the other objective based. In the policy-based approach, often dubbed supervised imitation learning, a policy function contingent on the state variables is trained directly from observed data demonstrated by experts. The obvious advantage of this type of approaches is its intuitiveness and simplicity, as it typically involves no micro-level decision-making model. But this simplicity is also its Achilles heel, as the trained policy function often has hard time to generalize into new environments or contexts. In the attempt to overcome this shortcoming, an objective-based approach, deemed inverse reinforcement learning (IRL), similar to system identification, aims to derive a micro-level decision-making model that can re-produce or fit the observed behavioral data well. With the help of the rigidity of a micro-level decision-making model, generalization to new environments or contexts are commonly believed easier. Since driving scenarios in reality can vary enormously, generalizability should be regarded critically important. Hence, we will follow the objective-based learning approach in this paper.

The literature on identifying dynamic systems started as soon as when the original Kalman filter was proposed \cite{Kalman64} in the context of linear-quadratic control. More recent work along this line of inquiry in more general context of MDPs, also known as inverse optimal control, can be found in \cite{Krishnamurthy2010}. Since the work of Ng and Russell \cite{Ng2000}, the research area of inverse reinforcement learning has undergone an explosive growth, see \cite{Abbeel2004} for apprenticeship learning, \cite{Ramachandran2007} for Bayesian based IRL, \cite{Ziebart2008} for maximum entropy based approach, and \cite{Sharma2017} for conditional choice probability based IRL, all in the MDPs context. More recently, generalization to various new contexts can be found, such as in multi-agent or game theory settings \cite{Ziebart2010, Schwarting2019}, with multiple intentions \cite{Babes2011}. There is also a separate but parallel strand of literature in econometrics that are methodologically very closely related, see \cite{Aguirregabiria2010, Arcidiacono2011}. A common feature of the above work is that the inference emphasis is mainly aimed at estimating the relative weights of the features or components in the reward function. Furthermore, most of these require relatively heavy computation to solve repeatedly dynamic programming problems in one form or another. 

In contrast with majority of the IRL literature that concentrate more on inverse methodologies, our focus is on decision-making modeling itself. This is based on two considerations. The first consideration is more philosophical, and has something to do with human's bounded rationality. It is our belief that no realistic drivers can actually solve extremely complex optimization problems in real time demanded by either MDP or game theory based frameworks. Instead, human driving should be mostly relying on heuristics. The second consideration is more practical, and has something to do with computation burden. Typically, models based on MDP or game theory are very hard to invert, as they involve dynamic optimizations or solving Nash equilibria. If we are successful in finding a simplified and yet realistic human driving decision-making framework, it would be reasonable to expect that the two aforementioned predicaments can be overcome at the same time. Leveraging on a recently proposed computational framework in \cite{Dai2020}, we show in this paper that such a human driving decision-making model can indeed be found. Using human driving trajectory data recorded from Sugiyama experiment, which are naturalistic though in a narrow context, we show how the inversion can be done explicitly, deploying the standard state space model in which only static decoupled optimizations are needed, even though the context involves multiple interacting agents in a dynamic setting. 

In the specific area of modeling human driving behaviors by applying various forms of IRL, there exist already a large number of papers. It is impossible to review the entire literature in this area. We only mention those work that explicitly involve continuous states and actions, which we regard as necessary for modeling realistic human driving behaviors. One popular approach is based on the work of Levine and Koltun \cite{Levine2012} for continuous inverse optimal control with locally optimal examples. Work along this line can be found in \cite{Sun2018a, Sun2018b, Naumann2020}. The local optimality assumption affords a dramatic computational simplification, as evaluation of the likelihood function no longer requires solving dynamic programming problems repeatedly. However, theoretically, assuming expert's demonstration being optimal, even locally, is unconventional. More commonly accepted approach is to assume that expert's demonstrated behaviors are centered around the optimal with errors. The other popular approach is to deploy some form of approximation  to the maximum entropy method \cite{Ziebart2008}. In \cite{Wu2020, Rosbach2019, Huang2020} sampling based approaches were used to estimate the partition function in the likelihood function. In \cite{Kuderer2015} the expected feature vector is approximately evaluated only on the most likely trajectory, a method also known as inverse optimal control. Yet another popular approach is to utilize techniques from deep learning literature. Work along this line can be found in \cite{Wang2020} who used semantically augmented adversarial IRL, and in \cite{Rosbach2020} who used attention networks for driving context switching. Some of these work use human driving of prototype vehicles on specialized road courses to provide expert demonstration data, while others use driving simulators. Neither of these can be hardly viewed as naturalistic. Driver heterogeneity and interaction among road users were explicitly handled in some, but not in others. However, ego vehicle and environmental vehicles are always treated asymmetrically. Another common characteristic is that they only estimate either the weights in a linear reward function or a neural network represented reward that is opaque. Furthermore, none of these made any attempt to address issues related to collective behaviors.

Methodologically, one paper \cite{Menner2020} that apparently appears close to our work, because it also culminates in a likelihood function akin to Kalman filter, derived from Bayesian perspective. However, their model is completely formulated differently from ours. For example, they did not propose a driving decision-making model explicitly. Instead, they treated driving dynamics as constraints enforced through measurement equations. In so doing, there is no reward function to learn. The priority tradeoff that were typically modeled using different weights in IRL was indirectly handled by the degree of the uncertainty in the corresponding measurement equation. An immediate consequence is that their training data have to include the complete states and actions, only possible in specialized environments, such as in a simulator like CarSim. In contrast, in our formulation Kalman filter is used naturally as the state evolution, while the human driving behavior is inserted into the filter through the standard control input via a micro level driving decision-making model. As a result, we only need the true measurement of vehicle's trajectory. Such a measurement process can be easily fulfilled naturalistically with external equipments, such as devices utilized by Sugiyama experiment \cite{Sugiyama08, Tadaki13, Wu2017, Stern2018} or by using aerial videos \cite{NGSIM, HighD, INTERACTION}.

As we will see later our modeling framework, once expressed in a form of state space model in Eq.(\ref{ssm_form}), can be viewed as a discretized version of the intelligent driver model with stochasticity recently introduced by Treiber and Kesting \cite{Treiber2017}, which unifies many popular car-following models into a single framework, including the optimal velocity model \cite{Nakayama2016}. The main differences are two: 1) we replaced their acceleration function, also known as optimal velocity function, by a behavioral model via utility maximization contingent on a general state variable; and 2) the effects of deterministic time delay and inertial/memory or ``action-stickiness'' are now represented by an auto-regressive process. One advantage of our approach is its ability to model driving behaviors with high fidelity and its flexibility to be generalizable to describing behaviors beyond simple car-following. Better generalizability is due to the fact that our approach is objective based once the utility function is calibrated, rather than policy based parameterized by the acceleration function in \cite{Treiber2017}. From the work in progress we will show how our model can be easily applied to urban settings, such as roundabout and intersection \cite{Wang2021, Song2021}, and to highway settings with or without ramps but also including lane changes \cite{Shen2021}. Another advantage is that the inverse problem of the forward decision-making model, i.e. estimation of model parameters, can be carried out using micro level vehicle trajectory individually by deploying the standard state space model technique.

To distinguish from the prior work in similar area, we specifically set the following simultaneous goals for our work in this paper: 
\begin{itemize}
\item The decision-making model is theoretically sound and intuitively interpretable, and its inversion involves relatively low computational demand; 
\item The human driving data have to be collected from non-specialized vehicles in a scalable manner so that they can be sufficiently naturalistic in a variety of realistic settings;
\item All agents are treated symmetrically, in the sense that ego and environmental agents are only for verbal description convenience, i.e. each agent can be simultaneously either ego or environmental;
\item The experimentally observed data can be fit well with all details at individual vehicle/driver level in a robust manner;
\item The estimated model parameters are meaningful and in the right range intuitively, so that they can also be used for other purposes, such as classifying driver types;
\item With the inferred parameters and noise level, the estimated model can produce via simulation non-trivial collective behaviors, such as spontaneous formation of stop-and-go shockwaves. Main features, such as queue length, average speed, speed range/variation, are re-produced almost quantitatively without accidents;
\item The estimated model will enable us to derive insights, such as necessary ingredients for shockwave formation, and ask/answer a number of what-if questions;
\item The functional form of the utility function shows the potential to be generalizable to other settings that are slightly different, such as Tadaki setting \cite{Tadaki13} and single-lane highway.
\end{itemize}

To further simplify the task, we explicitly parameterize all the necessary functional forms so that our inference is essentially parametric. Due to the unavoidable redundancy in the model specification, the detailed functional forms are not particularly critical to achieve our goals, so long as they are mutually consistent. Being parametric makes our approach completely transparent and interpretable, in the sense that all decision-making process can be traced exactly with their proper rationalization once the noise processes are switched off. We have further noticed empirically that inferring weights alone is not sufficient to closely describe human driving behaviors even for the simple setting of Sugiyama experiment. Hence, we are forced to also introduce and calibrate parameters that characterize the shape of the features or components in the reward function. Of course, we could have chosen more sophisticated representation of the reward function, such as using neural networks, as in \cite{Wang2020, Rosbach2020, Wulfmeier2016, Finn2016}. But that would require us to give up the interpretability and possibly generalizability, a compromise that we would rather not to make.

The reminder of the paper is organized as follows. In the next section, we briefly review the setting of the Sugiyama experiment \cite{Sugiyama08} and the recorded traffic data. Due to the roughness of the raw position data, we outline a simple smoothing scheme for deriving vehicle velocities used as the state variables perceived by the drivers at the scene. In section \ref{InferenceMethod} we introduce our inference method. This is accomplished by first recapitulating the forward model of human driving decision-making, and then casting it into a standard state space model framework, from which the inversion using Kalman filter with control input becomes obvious. We also address some of the technical issues associated with the inversion process in the same section. Due to the severe co-linearity of the model specification, we devote the entire section \ref{SimulationStudy} with extensive simulation studies to figure out a good strategy to reduce the redundancy or to identify the invertible subspace of the model. We then apply the developed methodology to Sugiyama data and present the calibration results at individual level in section \ref{CalibrationResults}. Here we also check the robustness of the results and classify driver types according to the inferred model parameters. In section \ref{InsightStudy} we use the calibrated model as a simulation device to gain insights or ask/answer a number of non-trivial what-if questions by varying the Sugiyama setting somewhat. We finally summarize and point out a few directions for future inquiry in the last section.

\section{\bf Sugiyama Experiment and the Data}
In this section we briefly review the setup of Sugiyama experiment and the observed data. Details are referred to the original paper \cite{Sugiyama08}.

\subsection{\bf Experimental setting}
The physical setting of Sugiyama experiment consisted of $N=22$ vehicles driven by human drivers on a circular road with a circumference of 230 meters. A 360-degree video camera situated at the center of the circle served as measurement device. The final trajectory data correspond to a time period length of $\Delta t=1/3$ seconds. Unfortunately, Sugiyama experiment did not record vehicle orientation. Consequently we will not be able to calibrate utility components and parameters in lateral direction in this study. The drivers were instructed to follow the vehicle ahead in safety in addition to trying to maintain their cruising velocity target of 30 km/h (or 8.3 m/s). This in turn implies that drivers have the tendency to chase the vehicle ahead when it is safe to do so.

\subsection{\bf Vehicle sizes}\label{Vehicle_size}
We measured vehicle longitudinal sizes using the video by counting the number of pixels for each vehicle. One pixel roughly corresponds to a length of 0.2 m, which gives us a sense of the accuracy on our estimation of vehicle sizes. The results are displayed in Table.\ref{VehicleSize}. It is noteworthy that the vehicle size heterogeneity is substantial, with smallest one barely at 3 meters and largest ones close to 5 meters.
\begin{table}[!h]
\begin{center}
\caption{Vehicle sizes (m) measured from the video.}
\begin{tabular}{l|c|c|c|c|c|c|c|c|c|c|c}
\hline
Vehicle \# & 0 & 1 & 2 & 3 & 4 & 5 & 6 & 7 & 8 & 9 & 10 \\
Size (m) &3.52 & 4.65 & 3.68 & 4.54 & 3.90 & 4.92 & 4.87 & 5.03 & 4.62 & 4.81 & 4.92 \\
\hline
\hline
Vehicle \# & 11 & 12 & 13 & 14 & 15 & 16 & 17 & 18 & 19 & 20 & 21 \\
Size (m) & 3.95 & 4.11 & 3.95 & 4.49 & 4.81 & 4.81 & 3.57 & 4.38 & 4.06 & 3.08 & 4.71 \\
\hline
\end{tabular}
\label{VehicleSize}
\end{center}
\end{table}

\subsection{\bf Longitudinal speed and acceleration}\label{EstimatedState}
The raw position data for vehicle $i$ at time $t$ denoted by $z_{i,t}^{(0)}$ was recorded by the central measurement system. It was the azimuth angle of the vehicle relative to the center of the circle that was actually measured. Due to the large measurement errors in the raw data we perform the following data smoothing progressively:
\begin{equation}
z^{(k)}_{i,t} = \frac{1}{4} \Big( z^{(k-1)}_{i,t-1} + 2z^{(k-1)}_{i,t} + z^{(k-1)}_{i,t+1}  \Big)
\end{equation}

The naive $k$th-smoothed velocity $\{\tilde{v}^{(k)}_{i,t}\}$ and acceleration $\{\tilde{a}^{(k)}_{i,t}\}$  of each vehicle can then be obtained as:
\begin{equation}
\tilde{v}^{(k)}_{i,t} = \frac{z^{(k)}_{i,t+1} - z^{(k)}_{i,t}}{\Delta t}\, \quad\text{and}\quad
\tilde{a}^{(k)}_{i,t} = \frac{ \tilde{v}^{(k)}_{i,t+1} - \tilde{v}^{(k)}_{i,t}}{ \Delta t }\, .\label{SmoothedState}
\end{equation}
In the parameter estimation process we treat $\tilde{s}_{i,t}=\{z_{i-1,t}^{(1)},\tilde{v}_{i-1,t}^{(2)},z_{i,t}^{(1)},\tilde{v}_{i,t}^{(2)},z_{i+1,t}^{(1)},\tilde{v}_{i+1,t}^{(2)}\}$ as driver $i$'s observed state at the scene at time $t$.

\subsection{\bf Vehicle trajectory and system-level speed profile}
\begin{figure}[!h]
\centering
\includegraphics[width=.80\textwidth]{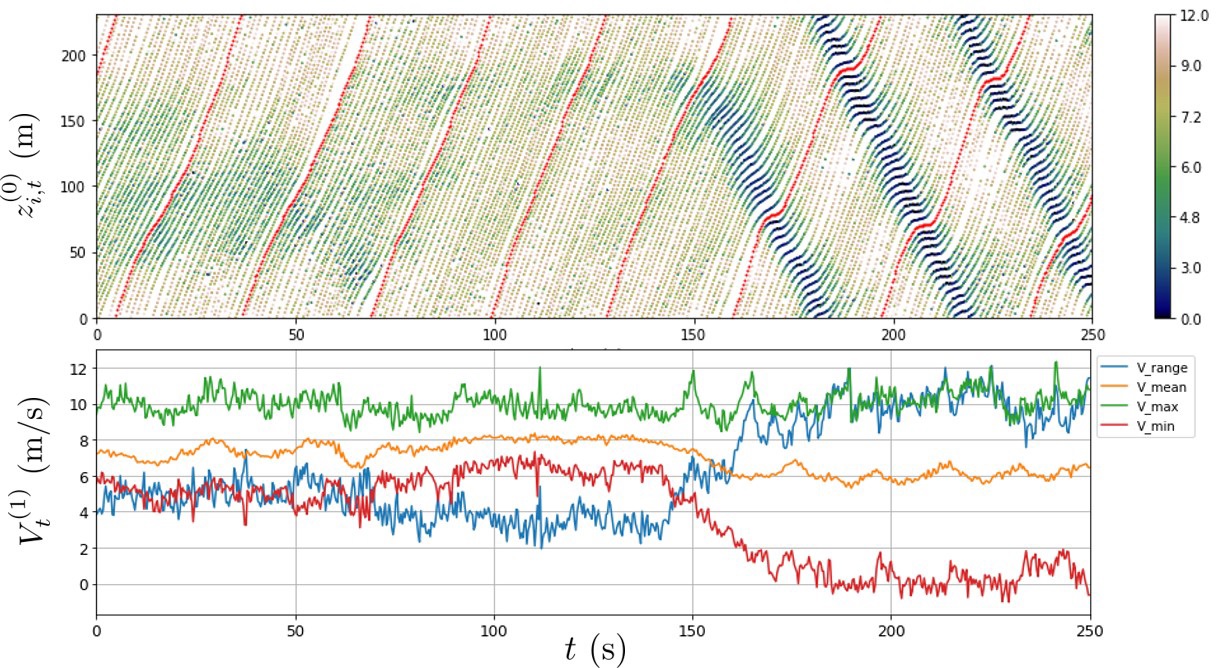}
\caption{Sugiyama experiment raw data: vehicle trajectories $z_{i,t}^{(0)}$ (upper panel) and system-level speed profile (lower panel) estimated based on once smoothed trajectory data $z_{i,t}^{(1)}$. The red curve in the upper panel is associated with the trajectory of vehicle $\#$0.}
\label{SugiyamaRawData}
\end{figure}

One of the most distinct features of the observed collective traffic pattern was the stop-and-go shockwave, which spontaneously forms without any bottlenecks or external disturbances, see Fig.\ref{SugiyamaRawData} for vehicle trajectories and system-level speed profiles. The system-level speed characteristics are defined as follows:
\begin{equation}
V_\text{avg}(t)=\frac{1}{N}\sum_i v_{i,t}\, ;\,\,\,
V_\text{min}(t)=\min_i\{v_{i,t}\}\, ;\,\,\, V_\text{max}(t)=\max_i\{v_{i,t}\}\, ;\,\,\, V_\text{range}(t)=V_\text{max}(t)-V_\text{min}(t)\, .
\end{equation}
The other important collective characteristics are the frequency and queue length of the slow moving vehicles and the backward traveling speed of shockwaves, which can be readily read off from the figure.

It is important to recognize that, while the experimental setting is not naturalistic as it can be, in addition to somewhat limited measurement accuracy, there are two main advantages for the recorded data that are quite unique: 1) each vehicle/driver was observed for a consecutive of 250 seconds along with all the necessary environmental (or state) variables; and 2) each vehicle/driver experienced enough state/action variations, such as episodes of fast and slow drivings, acceleration and braking cycles, periods of being far from and close to the vehicle ahead, and so forth. Necessary for the inference problem at individual level, these properties may not be always available in other more realistic settings. For example, a single drone captured video can only record consecutive driving for typically less than 60 seconds for each vehicle, which may not afford sufficient opportunity to sample the entire state/action space.

\section{Inference Method}\label{InferenceMethod}
In this section we outline the inference method, which consists of three parts: the forward problem with the explicit specification of the utility function, the state evolution, and inversion via maximum likelihood. We show how to put all these three parts together to form a standard state space model and its inverse. We conclude this section by mentioning how to tackle some of the technical issues in the inversion process.

\subsection{The forward problem: from utility to action}
In \cite{Dai2020} we proposed a heuristics based path planning algorithm called {\it adaptiveSeek}. In this approach Markov game with continuous actions and simultaneous moves was used as the modeling framework. The solution concept {\it adaptiveSeek} was simplified from a rigorous game theory based approach dubbed {\it betaNash} \cite{Dai2020} by relaxing some of the theoretical requirements that are too restrictive, both conceptually and computationally. We were able to show that {\it adaptiveSeek} can approximate the sub-game Nash equilibrium solutions well while taking into account the bounded rationality, at least in some simplified settings. We also used the same algorithm to simulate a number of more realistic cases and obtained very encouraging and intuitive results \cite{Wang2021, Song2021, Shen2021}. Therefore, we will regard {\it adaptiveSeek} as the forward decision-making model for human driving behaviors, from which we attempt to infer model parameters by inverting the decision process. 

Let's first recapitulate the decision-making algorithm {\it adaptiveSeek}, whose guiding principle is to follow human driving heuristics. The starting point is the simplified state evolution
\begin{equation}
{s}_{i,t+1}={f}_{i,t}(a_{i,t}|s_{i,t})\, . \label{StateEvolutionTilde}
\end{equation}
Typically, the utility function (per period) can be decomposed as a sum over several components:
\begin{equation}
u_{i,t}(a_{i,t}|s_{i,t},\kappa_i)=\sum_k \,\omega_i^{(k)}\,\phi_{i,t}^{(k)}(a_{i,t}|s_{i,t};\kappa_i)\, .
\end{equation}
In order to relax the strong assumptions associated with Nash equilibrium in {\it betaNash}, we have dropped, in both of these equations, the dependence on actions for all other agents by introducing an anticipation process in which we assume $a_{-i,t}$ has its natural extension. In the simple Sugiyama setting this amounts to assuming $a_{-i,t}=0$. However, whenever it is possible to do so, we will keep model parameters, such as utility parameter $\kappa_i$ and those in the state evolution, to be individual vehicle/driver dependent so as to keep track of the heterogeneity explicitly.

Then, the ideal action in {\it adaptiveSeek} is derived for agent $i$ at time period $t$ by the following optimization
\begin{equation}
{a}_{i,t}^*(s_{i,t}|\kappa_i)=\argmax_{a_{i,t}} \tilde{u}_{i,t}(a_{i,t}|s_{i,t};\kappa_i,h)\, , \label{BestResponseTilde}
\end{equation}
where the effective utility function with $h$-period anticipation is defined as follows.
\begin{equation}
\tilde{u}_{i,t}(a_{i,t}|s_{i,t};\kappa_i,h)=\sum_{k}\, w_{i,k} \,g_k\Big(\phi_{i,t}^{(k)}(a_{i,t}|s_{i,t};\kappa_i);h\Big)\, .\label{EffectiveUtilityTilde}
\end{equation}
The choice of the functional form for $g_k()$ depends on the specific component of the utility function and is made conservatively when necessary for safety reasons. For some components such as the moving forward reward and lane departure penalty, $g_k()$ are the average of the corresponding components in $h$-period. For driving smoothness, we choose the penalty in the first period. For components that are potentially calamitous, such as crash or collision penalties, we choose maximum penalty among the $h$ periods given the prescribed action sequences for others in the anticipation. Numerically, we choose $h=4$, corresponding to a look-ahead time horizon of $h\,\Delta t=4/3$ seconds. Explicit formulas for $g_k()$ will be given in subsections \ref{Moving_forward} and \ref{Pairwise_collision}, where we specify the utility function.

\subsection{\bf Utility function and its parameterization}
The utility function in our context is generally a device for aiding decision-making, and hence its interpretation can be either physical or subjective related to reward shaping \cite{Ng1999}. It is also important to realize that there could be generally substitution between the anticipation process in adaptiveSeek and the specific functional form of the utility function. As to which way of implementing a desired mechanism is a matter of convenience. Our experience is more along the line of making anticipation more generic and utility function more specific to the context.

With respect to Sugiyama experiment, we will need a minimum of two components. The first is the moving forward reward at a desired speed, and the second interactions among agents: pairwise collision penalty. These two components are qualitatively consistent with the driving instruction provided to the drivers. With additional complexity for inference, other components can also be added, see for example \cite{Naumann2020}, such as lane-departure penalty, or acceleration/braking-harshness penalty, driving roughness penalty, and so on. However, given the limitation of the observed data derived from a slow driving setting, adding these components only marginally improve the fitting. Hence, we will simply take the minimal approach in this work. On the other hand, we find that it is not sufficient to just estimate weights, which appear linearly in the total utility function. The shape parameters for each component turn out to matter a lot in subtle ways, for both achieving good fit to the observed data and for re-producing collective behaviors properly in post estimation simulations. 

\subsubsection{Moving forward reward at a desired speed}\label{Moving_forward}
The first component represents the intention to move forward along the ideal path at a desired speed:
\begin{equation}
\phi_{i, t}^{(1)}=\bigg[1-\exp\Big(-10\big(v_{i,t}+0.25\big)\Big)\bigg]-\bigg[1-\exp\Big(-\Big(\frac{v_{i, t} - v_i^*}{\kappa_i^{(1)} v_i^*}\Big)^{2}\Big)\bigg]\, ,
\end{equation}
where $v_{i,t}$ is the longitudinal velocity of agent $i$ at time $t$, and $v_i^*$ the ideal speed. The first term also discourages strongly the vehicle moving backward. The parameter $\kappa_i^{(1)}$ controls the degree by which agent $i$ likes to be close to its ideal speed. By normalization convention, we choose $\omega_i^{(1)}=1$. 

The anticipation related transformation function for this first component is given by 
$$g_1\Big(\phi_{i,t}^{(1)}(a_{i,t}|s_{i,t};\kappa_i);h\Big)=\phi_{i,t+1}^{(1)}\, .$$
If we were purely concerned with the regression quality another possibility would be equally good, with minor differences in estimated parameter values, 
$$g_1\Big(\phi_{i,t}^{(1)}(a_{i,t}|s_{i,t};\kappa_i);h\Big)=\frac{1}{h}\sum_{\tau=1}^h\phi_{i,t+\tau}^{(1)}\, .$$
However, we found that the first choice is better in re-producing the collective phenomena associated with the shockwaves.

\subsubsection{Pairwise collision penalty}\label{Pairwise_collision}
The second component represents the reward shaping of avoiding one-on-one collision with other vehicles, a subjective risk perceived by the driver. Let $L_i$ denote the length of vehicle $i$ (see subsection \ref{Vehicle_size}), and let $\mathcal{F}(x)=\exp{( -x^2-2x )}$. For vehicle $i$'s front collision, we choose
\begin{equation}
\phi_{i, j, t}^{(2)}=\left\{
\begin{array}{ll}
1, & \Delta x_{i, j, t} \leq 0 \\ [6pt]
\mathcal{F} \left(\displaystyle\frac{\Delta x_{i, j, t}}{\sigma_{i,j, t}}\right), & 0<\Delta x_{i, j, t}
\end{array}\right. 
\label{PairwiseCollisionPenalty}
\end{equation}
where $\Delta x_{i,j,t}=(x_{j,t}-L_j/2)-(x_{i,t}+L_i/2)$ is the bumper-to-bumper distance of vehicle $i$ to vehicle $j$ who is immediately ahead of $i$. It is natural to choose the front scale parameter to be speed dependent: $\sigma_{i,j,t} = \kappa_{i,c}+\kappa_{i,v}|v_{i,t}| + \kappa_{i,d}\max\{v_{i,t}-v_{j,t},0\}$. The $\kappa_{i,d}$ term is needed for braking when the vehicle ahead is slower, as we assumed zero acceleration in the anticipation process. Because the discouragement for moving backward in the first term in $\phi_{i, t}^{(1)}$ we do not need to explicitly consider the rear collision penalty here.

The anticipation related transformation function for this second component is given by 
$$g_2\Big(\phi_{i,t}^{(2)}(a_{i,t}|s_{i,t};\kappa_i);h\Big)=\max\{\phi_{i,t+1}^{(2)}, ...,\phi_{i,t+h}^{(2)}\}\, .$$

\subsection{State evolution: kinematic particle model with control input}\label{ParticleModel}
Since we do not observe lateral information in this experiment, we will deploy the kinematic particle model. The latent state space for agent $i$ is defined by a 3-component vector $\xi_{i,t}=(x_{i,t},v_{i,t},a_{i,t})$. In the state evolution for agent $i$, the driver's action is modeled as a control input that is contingent on a state, $s_{i,t}$, perceived by the driver at the scene at time $t$,
\begin{equation}\label{ssm_form}
\begin{cases}
x_{i,t}&= x_{i,t-1}+\Delta t\,v_{i,t-1}  +\mu^x_{i,t}\,\,\,\pmod{C}\, ,\\
v_{i,t}&=v_{i,t-1}+\Delta t\, a_{i,t-1}+\mu^v_{i,t}\, , \\
a_{i,t} &=\rho\,a_{i,t-1}+ \Big(a_{i,t}^*(s_{i,t}|\kappa_i)- \rho\,a_{i,t-1}^*(s_{i,t-1}|\kappa_i) \Big)+\mu^a_{i,t}\, ,\\
a_{i,t}^*(s_t|\kappa_i)&=\arg\max_{a}\tilde u(a|s_{i,t};\kappa_i)\, .
\end{cases}
\end{equation}
In the above equation $\pmod C$ represents modular with respect to the circumference of the circular road. This is to take care of the fact that $x_{i,t}$ must obey a periodic condition. There are two reasons that motivate our introduction of an AR(1) process for the the acceleration noise: 1) human driving reflex is typically longer than $\Delta t=1/3$ (s) and vehicle system lag, hence we need an explicit mechanism for time delay; and 2) there are persistent structures in the smoothed acceleration data at a time scale of around 2 seconds, hence we need a mechanism to capture the ``action-stickness'' or the inertial/memory effect. The remaining noise processes for $(\mu_{i,t}^x$, $\mu_{i,t}^v$, $\mu_{i,t}^a)$ are assumed to be mutually uncorrelated IID (independent and identically distributed) normal processes. As mentioned in Introduction, Eq.(\ref{ssm_form}) can be viewed as a discrete version of the recently proposed intelligent driver model with stochasticity \cite{Treiber2017}. Our framework has more modeling flexibility because a general state dependence is built into the utility maximization via control input so that it can be applied to a variety of settings beyond just car-following.

The above state evolution is executed by all agents simultaneously step-by-step, as implied by the Markov game setting. In so doing, every agent is treated symmetrically, in the sense that each agent is simultaneously an ego agent in its own state evolution, but also appears as an environmental agent in other agents' state evolution, so long as they are within the scope of the relevant state variables.

For the inverse problem, it is important to conceptually distinguish the two state variables: $\xi_{i,t}$ and $s_{i,t}$. The former is perceived from the point of view of the system/data analyst, as if he/she is trying to re-construct where agent $i$ was, how fast it was moving and with what acceleration. In contrast, the latter is perceived by driver $i$ at the scene at time $t$, representing the information needed for his/her driving decision-making. Due to this distinction, $s_{i,t}$ should be treated as given and fixed when $\xi_{i,t}$ is being estimated by the system/data analyst. Of course, the system/data analyst does not know $s_{i,t}$, hence it needs to be estimated based on some appropriate proxy derivable from the raw data when the state space model is inverted. To accomplish that we use smoothed vehicles trajectories $\tilde{s}_{i,t}$, as mentioned in subsection \ref{EstimatedState}.

\subsection{The inverse problem: from action to utility}
The actual state $x_{i,t}$ is only observable up to an observation noise, which leads the following measurement equation
\begin{equation} 
z_{i,t} = x_{i,t}+\nu_{i,t}\, ,
\end{equation}
where the observation noise $\nu_{i,t}$ is also assumed to be an IID normal distribution. For simplicity we further assume that the measurement noise is independent from the state evolution noises. We can re-write the state evolution and measurement equation together to form a more compact state space representation:
\begin{equation} 
\begin{cases}
\xi_{i,t} &=\Phi_i\,\xi_{i,t-1}+\Upsilon\, c_{i,t}(s_{i,t}|\kappa_i)+\mu_{i,t} \\
z_{i,t} &=A\, \xi_{i,t}+\nu_{i,t} 
\end{cases}
\quad\quad\text{with}\quad
\xi_{i,t}=\begin{pmatrix}
x_{i,t} \\ v_{i,t} \\ a_{i,t}
\end{pmatrix}
\end{equation}
where
\begin{equation}
\Phi_i=\begin{pmatrix}
1 & \Delta t & 0 \\ 0 & 1 & \Delta t \\ 0 & 0 & \rho_i
\end{pmatrix},\quad
A=\begin{pmatrix}
1 & 0 & 0
\end{pmatrix},\quad
\Upsilon=\begin{pmatrix}
0 & 0 & 0 \\ 0 & 0 & 0 \\ 0 & 0 & 1
\end{pmatrix}, \quad
\Omega_{i,\mu}=\begin{pmatrix}
\sigma_{i,x}^2 & 0& 0\\ 0 & \sigma_{i,v}^2 & 0 \\ 0 & 0 & \sigma_{i,a}^2
\end{pmatrix}\, .
\end{equation}
The control input vector is given by
\begin{equation}
c_{i,t}(s_{i,t}|\kappa_i)=\begin{pmatrix}
0 \\ 0 \\ a_{i,t}^*(s_{i,t}|\kappa_i)-\rho_i\, a_{i,t-1}^*(s_{i,t-1}|\kappa_i)
\end{pmatrix}\, .
\end{equation}
The log-likelihood function can be derived by following the standard Kalman filter approach, see e.g. \cite{TimeSeriesBook},
\begin{equation}
\mathcal{L}_i(\theta_i) =-\frac{1}{2}
\sum_{t=1}^T \ln|\Sigma_{i,t}|-\frac{1}{2}\sum_{t=1}^T\epsilon_{i,t}'\,\Sigma_{i,t}^{-1}\,\epsilon_{i,t}\, , \label{SSM_Likelihood}
\end{equation}
where $\epsilon_{i,t}\equiv z_{i,t}-A\,\mathbb{E}[\xi_{i,t}|z_{i,t-1}]$ and $\Sigma_{i,t}\equiv \text{Cov}(\epsilon_{i,t})$. The unknown model parameters to be estimated are given by $\theta_i=\{\kappa_i, \rho_i,\Omega_{i,\mu},\sigma^2_{i,\nu} \}$.

There are three distinct advantages of using {\it adaptiveSeek} as the decision-making model. The first is that $\mathcal{L}_i(\theta_i)$ only involves essentially static optimizations, even we are dealing with an intrinsic dynamic setting. In some sense, the dynamics is implicitly embodied in the sequence of state variables $s_{i,t}$. This property makes the estimation much simpler than those used in the standard MDP-based IRL approaches, see \cite{Ng2000, Abbeel2004, Ziebart2008}, where some form of dynamic programming problems have to be solved. The second is that $\mathcal{L}_i(\theta_i)$ contains no actions of any other agents given $s_{i,t}$, as if each individual agent is strategically independent from one another. These drastic simplifications become possible in our approach because we were able to decompose the would-be coupled dynamic planning problem (or solving dynamic Nash equilibrium, see \cite{Schwarting2019}) into a sequence of independent optimization problems contingent on the observable state and anticipation in Eq.(\ref{BestResponseTilde}). The third is that our method uses directly observed trajectory data, as aerial videos would naturally imply. In this way we avoid using acceleration data, which typically are much more noisy and require certain ad hoc smoothing that can potentially bias the system.

\subsection{Some technical aspects}
To successfully find the maximum likelihood solution, a few technical aspects still need to be addressed explicitly.
\subsubsection{Making the likelihood smooth}
Typically, the optimal action defined in Eq.(\ref{BestResponseTilde}) is done by using a grid search for computational convenience. This in turn implies that the likelihood functions in Eq.(\ref{SSM_Likelihood}) are generally not smooth in model parameters $\theta_i$. Consequently, gradient based methods cannot be directly applied for maximizing the likelihood. To remove this non-smoothness, we replace ${a}^*_{i,t}(s_{i,t}|\theta_i)$ in Eq.(\ref{SSM_Likelihood}) by the mean action defined by
\begin{equation}
\bar{a}_{i,t}(s_{i,t}|\theta_i)= \sum_{a} \, a\,P_{i,t}(a|\tilde{s}_{i,t};\theta_i)\, . \label{MeanAction}
\end{equation}
We choose Boltzmann distribution for the action distribution,
\begin{equation}
P_{i,t}(a|\tilde{s}_{i,t};\theta_i)=\frac{\exp\big[ \lambda\,\tilde{u}_{i,t}(a|\tilde{s}_{i,t};h,\theta_i)\big]}{Z_i(\theta_i|\tilde{s}_{i,t})}\, , \label{MaximumEntropy}
\end{equation}
where the denominator is the static partition function defined as
\begin{equation}
Z_{i,t}(\theta_i|\tilde{s}_{i,t})=\sum_{a}\,\exp\big[ \lambda\,\tilde{u}_{i,t}(a|\tilde{s}_{i,t};h,\theta_i)\big]\, , \label{PartitionFunction}
\end{equation}
Here $\lambda$ is purely regarded as a regularization parameter, in contrast with the maximum entropy method \cite{Ziebart2008} where $\lambda$ plays the role of characterizing the extend of the noise. Note that the weighting distribution is peaked at that particular action where $\tilde{u}_{i,t}$ reaches the maximum. The following relation implies that $\lambda$ can be interpreted as a rationality parameter for agent $i$,
\begin{equation}
\bar{a}_{i,t}(s_{i,t}|\theta_i)\rightarrow {a}^*_{i,t}(s_{i,t}|\theta_i)\quad\text{when $\lambda\to\infty$}\, ,
\end{equation}
when the maximum of the utility is non-degenerate.
After some experimentation the value of $\lambda=200$ is found to be quite adequate throughout the paper, given the optimization grid of 41 points on the interval of $a\in [-6,4]$ (m/s$^2$). 
%We also implement a similar softmax in $\tilde{u}(\cdot|\theta)$ defined in Eq.(\ref{EffectiveUtilityTilde}).

\subsubsection{Numerical optimization}
A variety of numerical algorithms can be used to optimize the smoothed likelihood function. Since the landscape of the likelihood function can be complicated, we will need those methods that are not entirely local. The standard {\it basinhopping} algorithm implemented in Python package SciPy appears to be reasonably effective. Stochastic gradient descent could be another possible choice.

\subsubsection{Dealing with co-linearity}
It turns out that there are certain redundancies in specifying the utility function. For example, higher weighting for the collision term at a given risk premium is roughly equivalent to a lower weighting for a larger risk premium, since both achieve something similar: to keep the distance from the vehicle ahead farther. This kind of ambiguities implies that we are likely to encounter severe co-linearity when the observed data quality is given. To alleviate this problem, we use extensively a simulated version of the model to figure out a good strategy (see section \ref{SimulationStudy} for details).

\subsubsection{Dealing with serial correlations}
The data used in the likelihood function Eq.(\ref{SSM_Likelihood}) for agent $i$ are supposed to be the location $z_{i,t}$ at all observed $t$'s. This seemingly indicates that we can have hundreds of data points even if we only observe the agent for a few minutes. However, due to the fact our observation is made at very short time interval, 1/3 seconds in our case, the state variation will be typically quite small from one time period to the next. Consequently, the observed data are likely to have strong serial correlation. It is well known that the inference needs to be done more carefully when the serial correlation is present in the data. One practical way to deal with this problem is to use simulation to mimic bootstrapping (again see section \ref{SimulationStudy} for details).

\section{Pre-Calibration Simulation Studies}\label{SimulationStudy}
There are several purposes for this section: 1) to show how the original model can be recovered and to what degree, provided there is no model misspecification and when signal-to-noise ratio is high; 2) to figure out a good strategy for dealing with the aforementioned co-linearity problem; 3) to get a sense of magnitude for the uncertainty associated with estimated parameters.

We adopt the following strategy for pre-calibration simulation studies. We started by experimenting many functional forms and parameter combinations for the utility, to make the the simulation results look as close to the observed data from Sugiyama experiment as possible. We then used the functional form to fit the real data and see if the inference outcome was reasonably meaningful and satisfactory. We iterated these steps till we reach some sort of convergence. The final choice of the utility function is as specified in the previous section. 

\subsection{Invertible subspace of the likelihood}
We knew from experience that the utility function and other model parameters are not always well separated. Degeneracies due to high correlation among parameters, such as $\omega^{(2)}_i$ and risk premium $\kappa_{i,v}$, and insensitive parameters, such as $\sigma_x$ and $\sigma_v$ (so long as they are in the reasonable range), will lead to ill-conditioned inverse problem. To understand the degree to which the forward decision model is practically invertible we examine the Hessian of the likelihood function in the neighborhood of its maximum point, a task easily achievable in simulation. It turns out that only four largest eigenvalues are consistently negative definite. All other smaller eigenvalues fluctuate quite a lot, especially when the real data are used. This indicates that we can at best invert the model meaningfully in a 4-dimension subspace. 

Again, after some experimentations and also for ease of interpretation, we choose the following subspace and their values shared by all vehicles: 
\begin{equation}
\theta=\{\sigma_\nu=0.263, \sigma_a=0.273, v^*=10.26,  \kappa_v=0.215\}. 
\label{SimulatedModelParameters}
\end{equation}
For the purpose of testing the proposed inverse methodology in the right neighborhood of the parameter space these values are chosen to be close to the mean of the individually estimated parameters, except for $\sigma_a$, which is drastically reduced so that we can draw an unambiguous conclusion. Other ``less important or nearly degenerate'' parameters are all fixed intuitively as follows: $\kappa^{(1)}=0.7$, $\sigma_x=0.05$ (m), $\sigma_v=0.1$ (m/s), $\rho=0.7$, $\kappa_c=0.4$ (m), $\kappa_d=1.0$ (s), and $\omega^{(2)}=-10$.

\subsection{Inference at individual level}
\begin{figure}[!h]
\centering
\includegraphics[width=.80\textwidth]{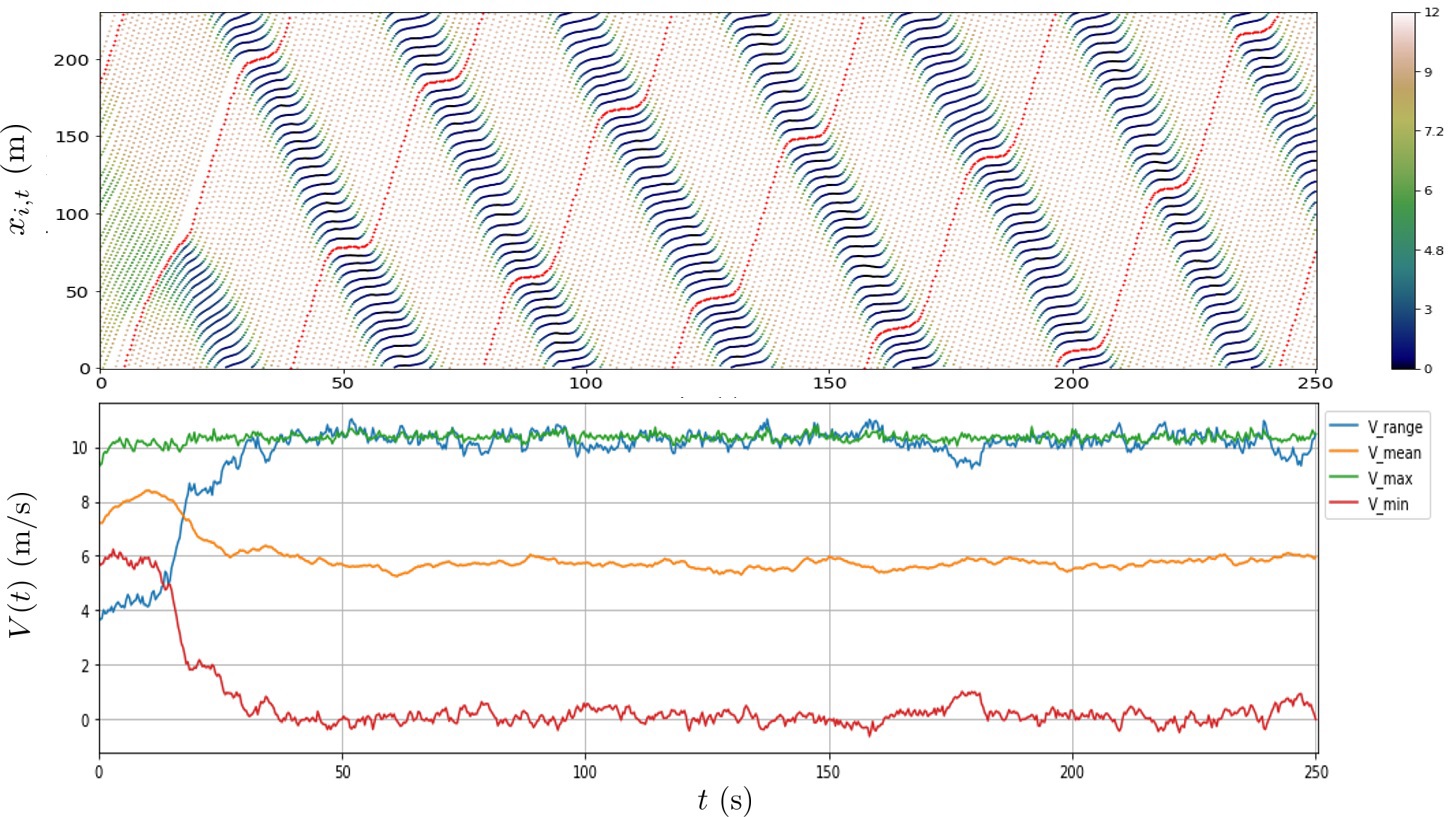}
\caption{The simulated vehicle trajectories (upper panel) and system-level speed profile (lower panel) using {\it adaptiveSeek} with the estimated utility parameters. The red curve in the upper panel is associated with the trajectory of vehicle $\#$0.}
\label{Simulated_PreCalibration}
\end{figure}

The simulated vehicle trajectories and  speed profile are shown in Fig.\ref{Simulated_PreCalibration}. Similarity with Sugiyama raw data in Fig.\ref{SugiyamaRawData} is quite evident. All major characteristics are re-produced well. The only exception is that the queue length is about one vehicle too long in the simulation.

\begin{figure}[!h]
\centering
\begin{tabular}{cc}
\includegraphics[width=.40\textwidth]{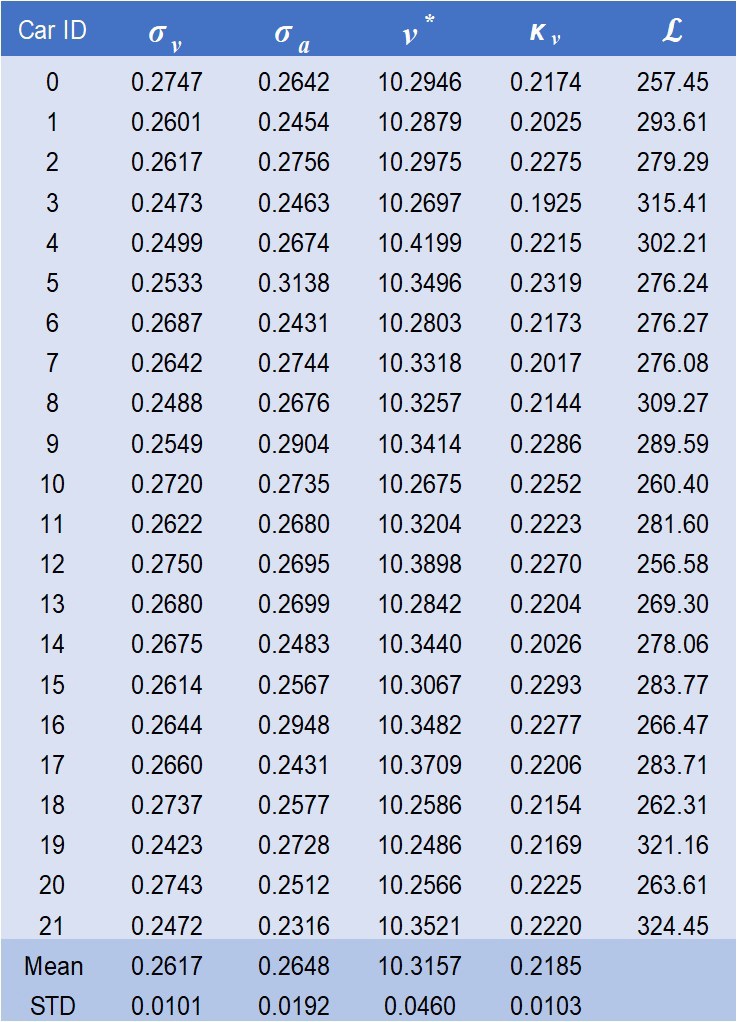}
\includegraphics[width=.46\textwidth]{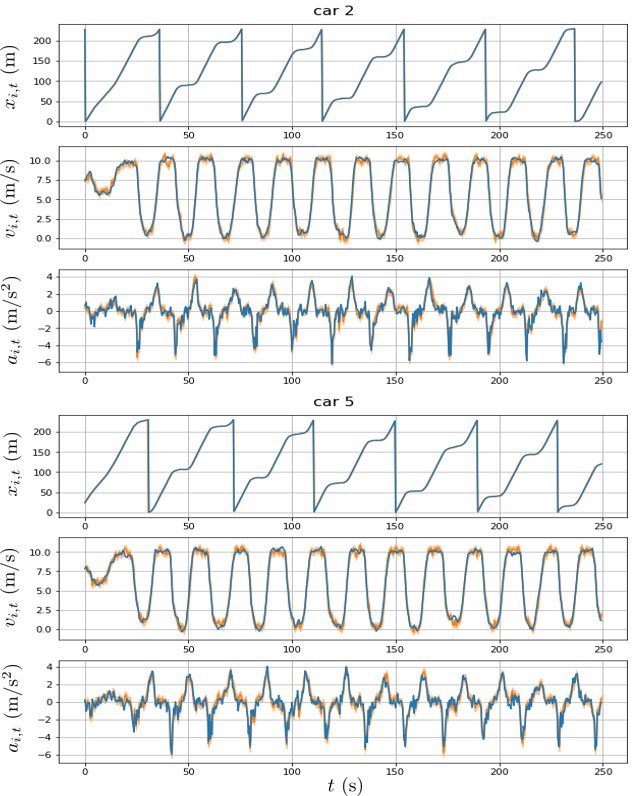}
\end{tabular}
\caption{(Left) Calibrated model parameters at individual level, with the standard deviation among all vehicles as a measure of heterogeneity; (Right) Typical in-sample fits where the blue curves are the simulated state variables (${x}_{i,t}$, ${v}_{i,t}$, ${a}_{i,t}$), and the orange curves are estimated latent state filters $\xi^t_{i,t}\equiv E[\xi_{i,t}|z_{i,t}]$. The shaded band around the orange curve represents one standard deviation extracted from $P_{i,t}^t\equiv \text{Cov}(\xi_{i,t})$.}
\label{Simulation_Results}
\end{figure}

The individual calibration results using the simulated data are summarized in Fig.\ref{Simulation_Results}, with estimated model parameters in the left penal and typical fits in the right penal. 
Not surprisingly, without systematic model misspecification and with a high level of signal-to-noise ratio, the estimated values for the four major model parameters are all centered around their respectively assigned values in Eq.(\ref{SimulatedModelParameters}). The in-sample fits also look very good, with all three state variables, location, speed and acceleration, well re-produced within the estimated error bands. This accurate recovering of the original simulation parameters and state vector gives us the confidence that our proposed SSM model approach to the inverse problem is on a solid foundation.

\subsection{Parameter uncertainty}
Because the likelihood function is close to be singular, the inverse of the Fisher information matrix cannot be used reliably as the measure of parameter errors. We also mentioned another potential issue in estimating parameter errors: the driver observed state variable, $s_{i,t}$, is highly correlated serially. Therefore, the only sensible way to obtain quantitative information on estimated parameter errors is to follow some sort of bootstrapping method, such as \cite{Bootstrapping1991}. However, since we are using simulation with known parameters for each vehicle that is a priori identical, we can get a good feel on the parameter estimation by simply examining the deviations of the regressed outcomes with the actual parameters across all the vehicles. When there is no model misspecification, we did not find any systematic bias for the four important parameters: $\{\sigma_\nu,\sigma_a,v^*,\kappa_v \}$. The corresponding magnitudes of the statistical errors are  0.01 m, 0.02 m/s$^2$, 0.05 m/s, and 0.01 s, respectively. 

\section{Calibration Results Using Sugiyama Data}\label{CalibrationResults}
Adhering to the strategy developed in the proceeding section, we fix following ``less sensitive or nearly degenerate'' parameters: $\kappa^{(1)}=0.7$, $\sigma_x=0.05$ (m), $\sigma_v=0.1$ (m/s), $\rho=0.7$, $\kappa_c=0.4$ (m), $\kappa_d=1.0$ (s), and $\omega^{(2)}=-10$. All these are chosen to be the same for every individual. The explicit inference parameters via likelihood maximization for agent $i$ are then given by: $\theta_i=\{\sigma_{i,\nu}, \sigma_{i,a}, v_i^*,  \kappa_{i,v}\}$.

\subsection{Additional regularizations}
Even limited within the 4-dimensional subspace it is still not possible to do the inverse at individual level by strictly following Eq.(\ref{SSM_Likelihood}) alone. This probably should not be too surprising, due to the unavoidable model misspecification and the relatively poor level of signal-to-noise ratio of the raw data. The problem mostly manifests in two aspects: some of the estimated ideal speed ($v^*_i$) being too high or the estimated risk premium ($\kappa^*_{i,v}$) being too low, for at least a few agents, though not necessarily at the same time. Being out of the reasonable range for these estimated parameter values, while affecting little to the quality of the in-sample fit, can in turn imply very different collective behaviors once all the vehicles are put back together for a post-calibration simulation. To soften this problem, we introduce the following regularizations in the ideal speed and risk premium subspace:
\begin{equation}
\mathcal{L}_i(\theta_i) \rightarrow \mathcal{L}_i(\theta_i) -\gamma_v \big(v_i^*-\bar{v}\big)^2
-\gamma_\kappa \big(\kappa_{i,v}^*-\bar{\kappa}_v\big)^2\, , \label{Regularized_SSM_Likelihood}
\end{equation}
where the common values for the ideal speed and risk premium are chosen as $\bar{v}=11.0$ (m/s) and $\bar{\kappa}_v=0.5$ (s). Even though it can be interpreted along the line of empirical Bayes enforcing the prior information on the collective behavior, adding regularization in this way still seems to be somewhat artificial. But the choice of their values are intuitively understood as follows. Regularization on the ideal speed is to pull down the estimated $v_i^*$ from reaching values as high as $14$ (m/s) for a few vehicles, which in turn would make formation of shockwave impossible, due to extremely high ideal speed heterogeneity. Likewise, if we were not regulating the risk premium by setting a higher $\bar{\kappa}_v$, a few estimated $\kappa_{i,v}^*$ could end up being too low, which in turn would imply collisions for subsequent simulations. Empirically, we found that the common values of $\gamma_v=50$ and $\gamma_\kappa=500$ are reasonable for the purpose of regularization for all agents, while the effect on in-sample fit and other parameters is minor. Numerically, these regularization terms are in the order of a few percent of the un-regularized likelihood.

\subsection{Inference at individual level}
\begin{figure}[!h]
\centering
\begin{tabular}{lr}
\includegraphics[width=.43\textwidth]{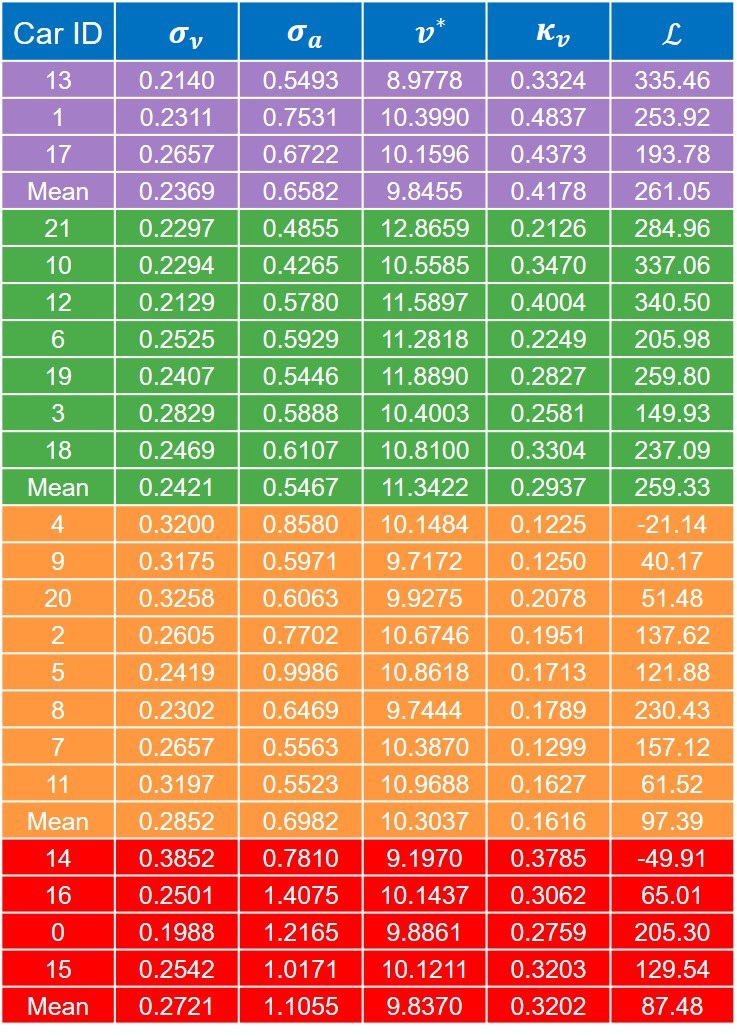}
\includegraphics[width=.48\textwidth]{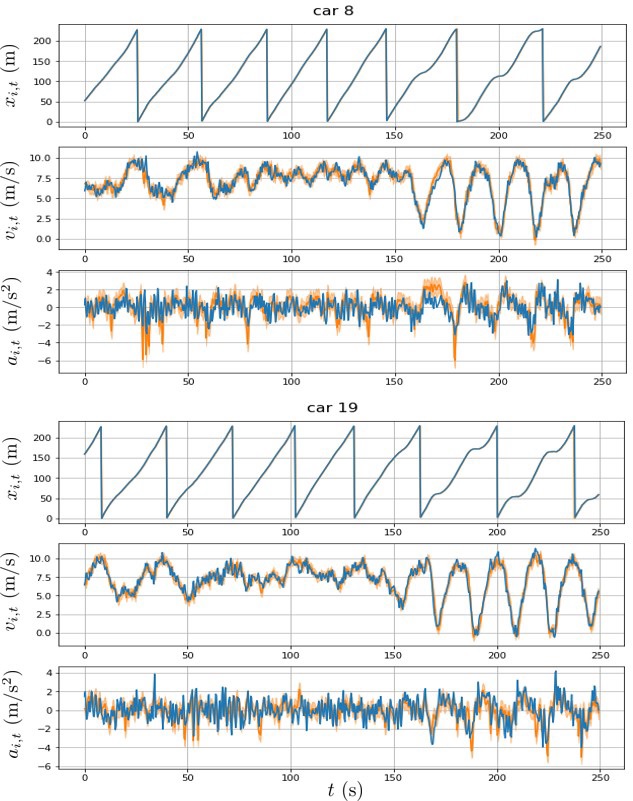}
\end{tabular}
\caption{(Left) Calibrated model parameters at individual level; (Right) Typical in-sample fits where the blue curves are naively smoothed speed ($\tilde{v}_{i,t}^{(1)}$) and acceleration ($\tilde{a}_{i,t}^{(2)}$) defined in Eq.\ref{SmoothedState}, and the orange curves are estimated latent state filters $\xi^t_{i,t}\equiv E[\xi_{i,t}|z_{i,t}]$. The shaded band around the orange curve represents one standard deviation extracted from $P_{i,t}^t\equiv \text{Cov}(\xi_{i,t})$.}
\label{SugiyamaCalibrationResults}
\end{figure}

Following Eq.(\ref{Regularized_SSM_Likelihood}) the calibration results are summarized in Fig.\ref{SugiyamaCalibrationResults}, with model parameters in the left penal and typical fits in the right penal. 
The reason for the unconventional ordering of the agents will become clear shortly in the next subsection. A few comments are in order here. First, the estimated model parameters all turn out to be in the reasonable range intuitively, except for the likely exaggeration of $\sigma_a$. Second, there are substantial heterogeneity among all agents. Third, the estimated measurement errors ($\sigma_\nu$) are in the same order of magnitude as the pixel size (or about 0.2 m). Fourth, while they are systematically higher than the advised cruising speed of 30 km/h by about 25\%, the estimated ideal speeds ($v^*$) are centered around the observed long-time mean of the maximum speed (c.f. Fig.\ref{SugiyamaRawData}). This in turn implies that drivers did not naively follow the driving instruction, but rather adapted to the specific driving condition changes. Fifth, the magnitude of the speed dependent risk premium coefficients ($\kappa_v$) implies a headway of at least $\sigma_{i,t}\sim 1.6$ m in Eq.(\ref{PairwiseCollisionPenalty}) at the average speed of $6$ m/s during the shockwave phase of the raw data. Sixth, comparing to the fits found in the simulation study in Fig.\ref{Simulation_Results}, where there is no model misspecification, the fits with real data are not as good, as expected. Though the speed is still well re-produced, the acceleration suffers much larger errors. Note that the blue curves in Fig.\ref{SugiyamaCalibrationResults} are not the ground  truth of the actually realized state variables, but only naively estimated ones using smoothing technique. Nevertheless, the main features of the action characteristics are captured properly during periods with and without shockwaves.

There are many possible sources of model misspecification. These can include any or all the following: 1) $\Delta t=1/3$ seconds is too large; 2) the utility is not complete or has slightly different functional form; 3) the data are too noisy, perhaps with unknown systematic observation noises; 4) the likely discrepancy between $\tilde{s}_{i,t}$ and the actually perceived state by drivers at the scene; 5) the ill-conditioned nature of the Hessian so that a lot of the ``minor parameters'' need to be fixed by hand without being able to build in the proper heterogeneity. Because of these possibilities, we believe that the action uncertainty ($\sigma_a$) is likely to be exaggerated substantially. 

Given the magnitudes of estimated $\sigma_a$ using real data are up to 5 times of that in the pre-calibration simulation, the statistical errors for $(\sigma_\nu, \sigma_a, v^*, \kappa_v)$ can be estimated by scaling the simulated pre-calibration results in Fig.\ref{Simulation_Results}, yielding $(0.05, 0.1, 0.25, 0.05)$ as a very rough order-of-magnitude estimation for the corresponding parameter errors in Fig.\ref{SugiyamaCalibrationResults}.

\subsection{Classification of driver types}
Given the inference results at individual level, we can categorize driver types using their estimated parameters. The reason that we favor using the estimated parameters for this purpose, rather than some chosen pattern of action sequences, is that the latter could be confounded by the traffic environment embodied in the state variables. This confounding is completely removed in the intrinsic space of model parameters, which depends only on the individual's driving preference given the context. 

\begin{figure}[!h]
\centering
\begin{tabular}{lr}
\includegraphics[width=.46\textwidth]{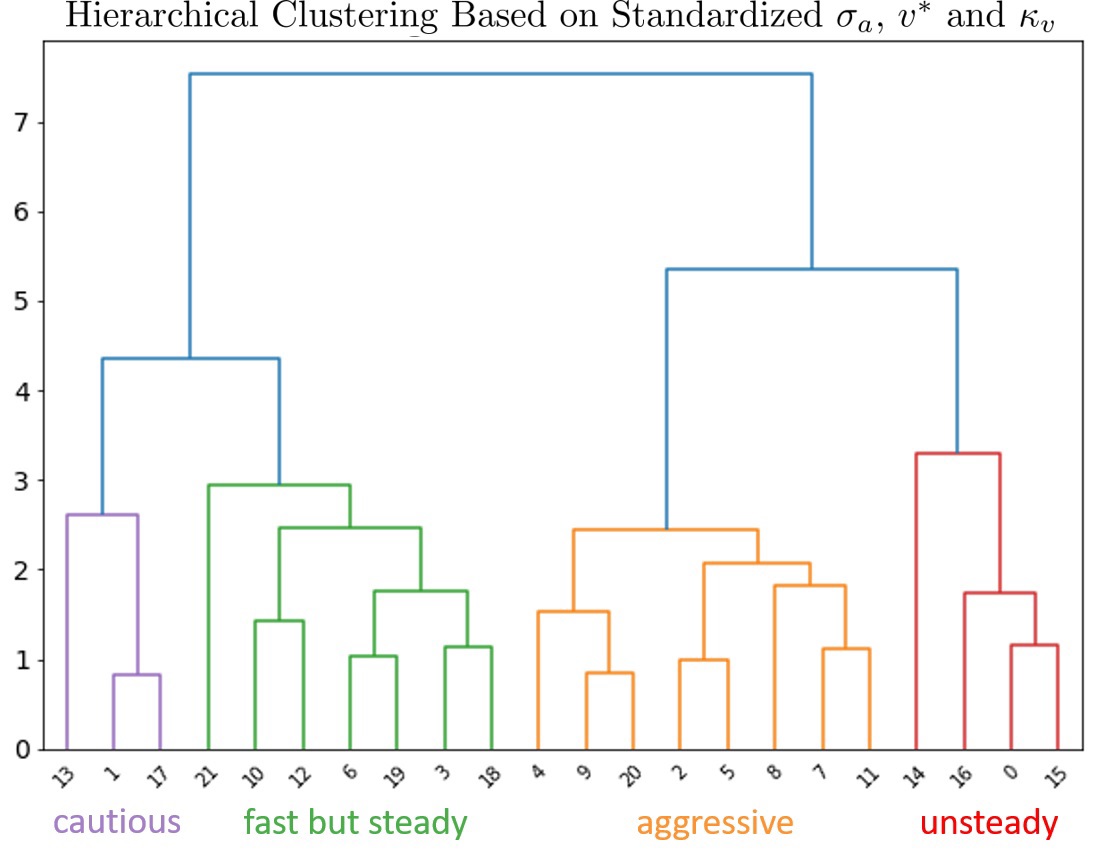}
\includegraphics[width=.46\textwidth]{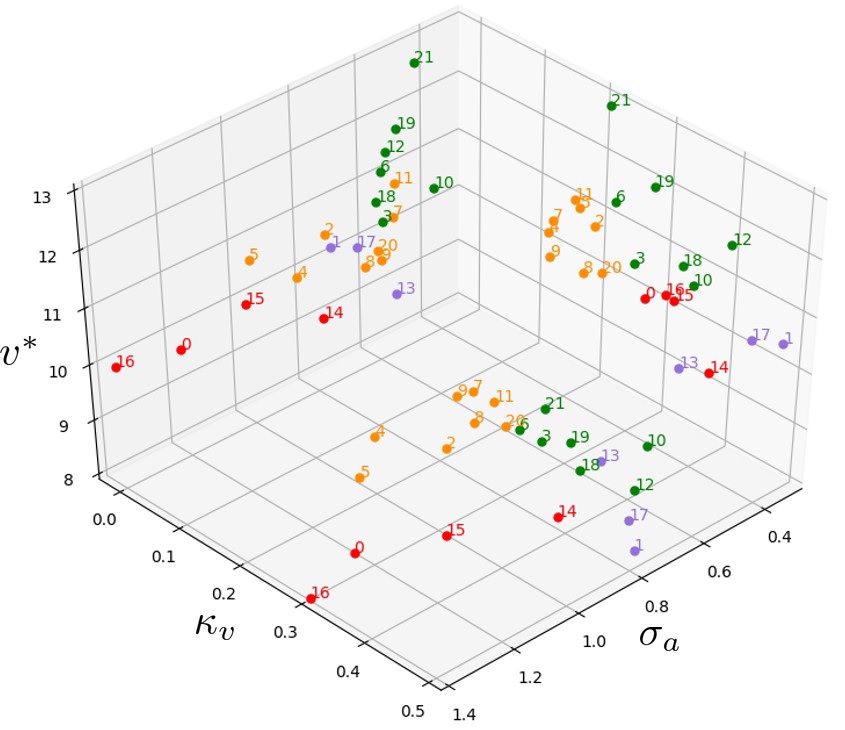}
\end{tabular}
\caption{Left: Driver type clustering based on the three individually estimated utility parameters (standardized): action uncertainty ($\sigma_a$), ideal speed ($v^*$), and speed-dependent risk premium ($\kappa_v$); Right: Projection of the estimated utility parameters onto the three planes $(\sigma_a,v^*)$, $(v^*,k_v)$, and $(k_v,\sigma_a)$, showing how these parameters are distributed and to what degree they are separated.}
\label{SugiyamaClustering}
\end{figure}

One idea to try is hierarchical clustering in the 3-dimensional model parameter space ($\sigma_a$, $v^*$, $\kappa_v$).\footnote{We exclude the measurement error $\sigma_\nu$ in the clustering, because it is mostly a system-level parameter unassociated with any individual vehicle/driver.} These three parameters capture three distinct driving characteristics: uncertainty of acceleration or braking (a measure of execution steadiness from the optimality), tendency to drive fast or slow when no other vehicles around (a measure of hastiness or urgency), and how much headway margin the driver is willing to keep to the vehicle ahead (a measure of risk/aggressive attitude). Hierarchical clustering method is preferred because we would like to have the number of clusters to emerge endogenously. Euclidean distance in the standardized parameter space is used as the measure for closeness. The clustering result is depicted in Fig.\ref{SugiyamaClustering}. Four broadly meaningful clusters can be identified, represented by four different colors. By looking at their group-level  parameter averages, we can make the following driver-type labeling, based on the most distinct characteristics of the group while ignoring the dimensions that are close to the total averages: cautious drivers (purple); fast but steady drivers (green); aggressive drivers (orange); and unsteady drivers (red).

Since we do not have a priori information on driver types, how do we know that the clustering outcome is meaningful? A nice way to validate our result is based on the observation that the average speed during the shockwave phase (roughly from 150 to 250 s) in Sugiyama raw data (see the orange line in Fig.\ref{SugiyamaRawData}) displays a quasi-sinusoidal pattern. It turns out that this pattern is associated with the bunch of vehicles ranging from $\#2$ to $\#11$ (mostly agents in aggressive class (orange) in Fig.\ref{SugiyamaCalibrationResults}) happen to be in the queue of the slow-moving vehicles, corresponding to the minimum points in the orange line. Our calibrated result provides a natural explanation: because of their low risk premium this bunch of vehicles prefer to closely chase the vehicle ahead, resulting in longer slow-moving queue and hence lower average speed in aggregate. To verify if this interpretation is truly relevant, we randomly re-shuffled the order of the vehicles and observed the disappearance of the sinusoidal pattern.\footnote{We actually observed this phenomenon in our pre-calibration simulation study. There is no such sinusoidal pattern in Fig.\ref{Simulated_PreCalibration} because we practically switch off the parameter heterogeneity there.}

\subsection{Robustness check of the estimated model}
To further validate our estimation results, we perform additional simulation experiments. The purpose is to check the robustness against meaningful but small variations to the precise condition at which the model is estimated. More specifically, the following situations are examined:
\begin{itemize}
\item Random shuffling the order of the vehicles: the quasi-periodicity of the average speed during the presence of shockwaves may disappear when $\kappa_{i,v}$ is more evenly distributed, while everything else remains more or less intact.
\item Making the time period shorter, i.e. $\Delta t=0.2$ (s): in this case, we also need to scale some of the time dependent variables, such as $h\rightarrow 6$ so that anticipation time window is still $h\times\Delta t=1.2$ (s), and $\rho\rightarrow\rho^{2/3}$. Then the simulated results look almost independent of the value of $\Delta t$.
\item Variation in circumference of the circle or vehicle density: spontaneous shockwave formation continued so long as the density does not become too low.
\item A number of other variations that aim to understand the formation of the shockwaves will be presented in the next section.
\end{itemize}

\section{Insight Study Using Estimated Model}\label{InsightStudy}
Armed with the estimated utility function and other model parameters for each agent, we can simulate a number of situations, partially to validate the calibration procedure, and partially to see if the calibrated model can be used to gain insights at slightly varied contexts. In addition to the estimated utility parameters, the likelihood method also provide an estimation of the noise level $\sigma_a$, which we believe is exaggerated, due to likely model misspecifications and high level of measurement noise. Given the AR(1) noise structure, the theoretically predicted action at time $t$ contingent on the state and the information known up to time $t$ is given by
\begin{equation}
a_{i,t}(s_{i,t})=\bar{a}_{i,t}(s_{i,t}|\kappa_i^*)+\rho_i\Big(a_{i,t-1}(s_{i,t-1})-\bar{a}_{i,t-1}(s_{i,t-1}|\kappa_i^*)\Big)
+\tilde{\epsilon}_{i,t}\, ,
\end{equation}
where $\kappa_i^*$ presents the utility parameters estimated via maximum likelihood, $\tilde{\epsilon}_{i,t}$ obeys IID $N(0,\tilde{\sigma}_{i,a}^2)$, with $\tilde{\sigma}_{i,a}=0.5\times\sigma_{i,a}$. The deflating factor of 0.5 is empirically chosen so as to prevent collisions in the simulation while leaving everything else intact as much as possible.\footnote{Even with reduced acceleration uncertainty by half, collisions could still happen in simulations with long time horizon, though extremely rare. It is our belief that the root of the problem is the assumption of $\sigma_{i,a}$ being constant. When the speed of the vehicle is low, it is perhaps more reasonable to assume that the acceleration uncertainty is proportional to the speed. Of course, this will make the estimation harder, a topic to be tackled in the future.}

\subsection{Simulated Sugiyama experiment}
\begin{figure}[!h]
\centering
\includegraphics[width=.80\textwidth]{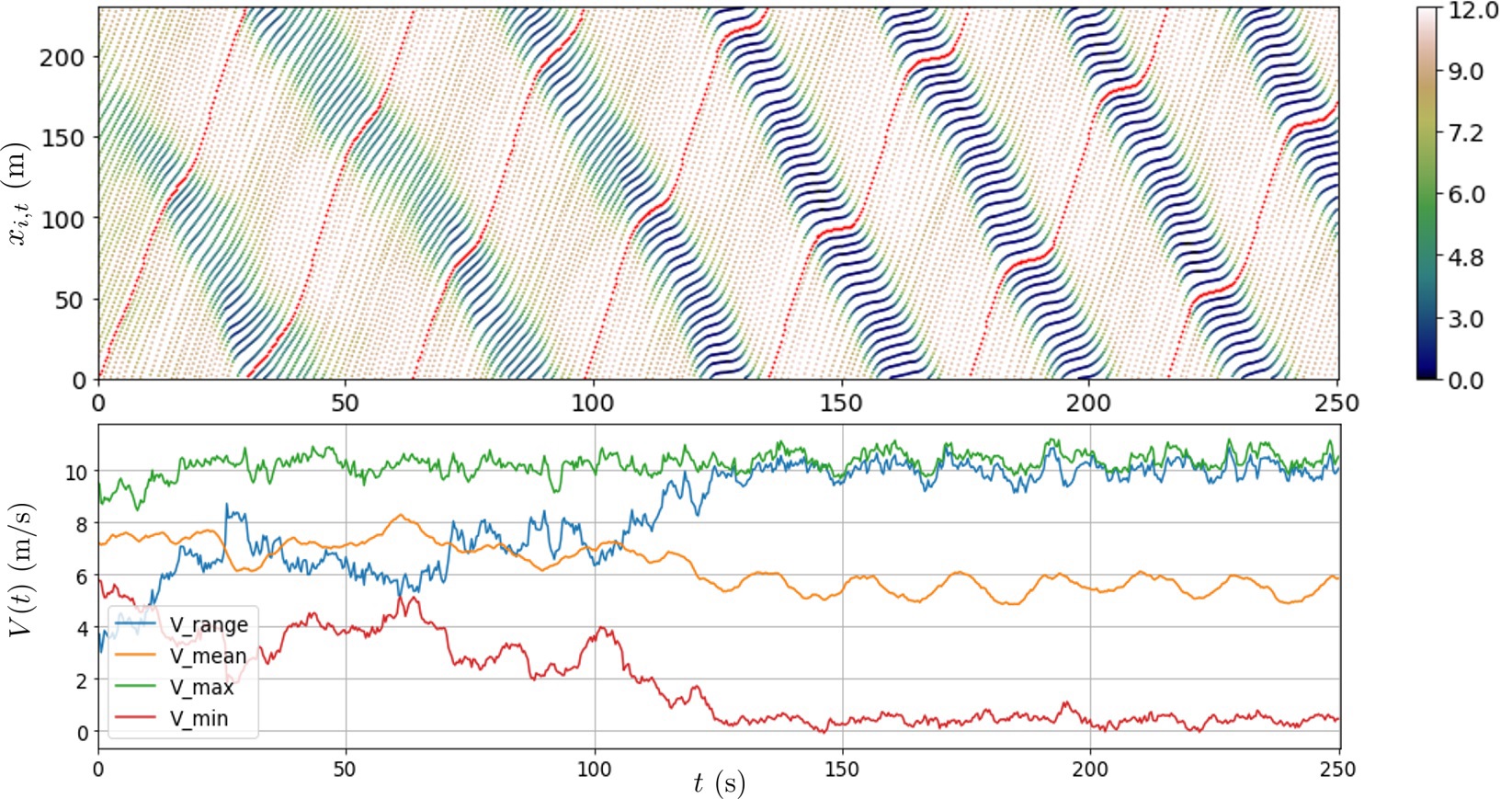}
\caption{The simulated vehicle trajectories (upper panel) and system-level system-level speed profile (lower panel) using {\it adaptiveSeek} with the individually estimated utility parameters. The red curve in the upper panel is associated with the trajectory of vehicle $\#$0.}
\label{SugiyamaSimulated}
\end{figure}

We can now do a direct post-calibration validation by simulating the forward problem using {\it adaptiveSeek} with the estimated parameters. An example of simulated vehicle trajectories and the system-level speed profile, using utility parameters estimated at individual level, can be found in Fig.\ref{SugiyamaSimulated}. The qualitative resemblance between this figure and Fig.\ref{SugiyamaRawData} is evident. It is important to recognize that, while the calibration is performed purely at individual level, as if each agent is not interacting with one another given the driver perceived state $\tilde{s}_{i,t}$, the simulated outcome displays proper collective phenomena, such as the spontaneous formation of stop-and-go shockwaves. Even some of the details, such as the minimum and maximum speeds, queue frequency, backward traveling speed of the shockwaves, and the quasi-sinusoidal pattern of the average speed, are all close to the observed data. The only exception is the average speed being about 0.5 (m/s) too low, due to the queue length being slightly longer (by about one vehicle). This type of agreements indirectly, therefore, validates our approach in a highly non-trivial manner.

\subsection{Insight studies}
Now we would like to vary the simulation setting somewhat so that we can derive a number of insights on what makes the formation of stop-and-go shockwaves possible or impossible.

\subsubsection{Heterogeneity and acceleration uncertainty}
\begin{figure}[!h]
\centering
\includegraphics[width=.70\textwidth]{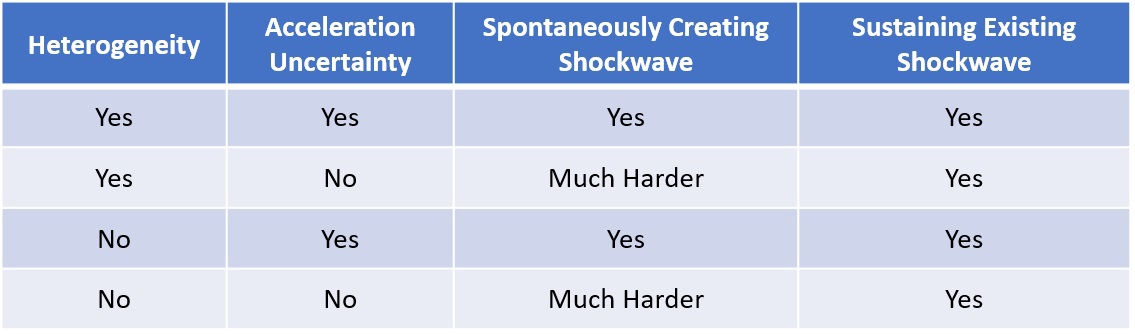}
\caption{Results of ablation studies: impact of switching on/off heterogeneity and acceleration uncertainty to spontaneous creating or sustaining existing stop-and-go shockwaves.}
\label{AblationResults}
\end{figure}

What if we switch off heterogeneity or acceleration uncertainty, either separately or at the same time. This is similar to ablation studies often found in reinforcement learning literature. To switch off the heterogeneity, we assign all agents to have the same parameters as the average agent. Switching off the acceleration uncertainty is achieved simply by setting $\sigma_{i,a}=0$. We are interested in seeing whether the shockwaves can be generated spontaneously or at least the shockwave can be sustained once it is already there. The simulation results are summarized in Fig.\ref{AblationResults}. Heterogeneity appears to be unnecessary for the formation of skockwaves or sustaining them. On the other hand, the acceleration uncertainty appears to be necessary for at least generating shockwaves spontaneously, though not necessary for sustaining existing ones. Our simulation results here is consistent with three common observations, as summarized in recent stochastic intelligent driver model \cite{Treiber2017}: 1) when studying traffic oscillations by using various car-following models heterogeneity plays minor role, as long as it is not too large; 2) the collective driving dynamics has to embody some form of intrinsic instabilities, and the sustaining capability of shockwaves indefinitely in the simulation suggests that {\it adaptiveSeek} has similar instability built in; and 3) acceleration uncertainties act as triggers to oscillation events, as there is no other external noise in the simulation process. 

\subsubsection{What if one vehicle has much lower ideal speed}
\begin{figure}[!h]
\centering
\includegraphics[width=.80\textwidth]{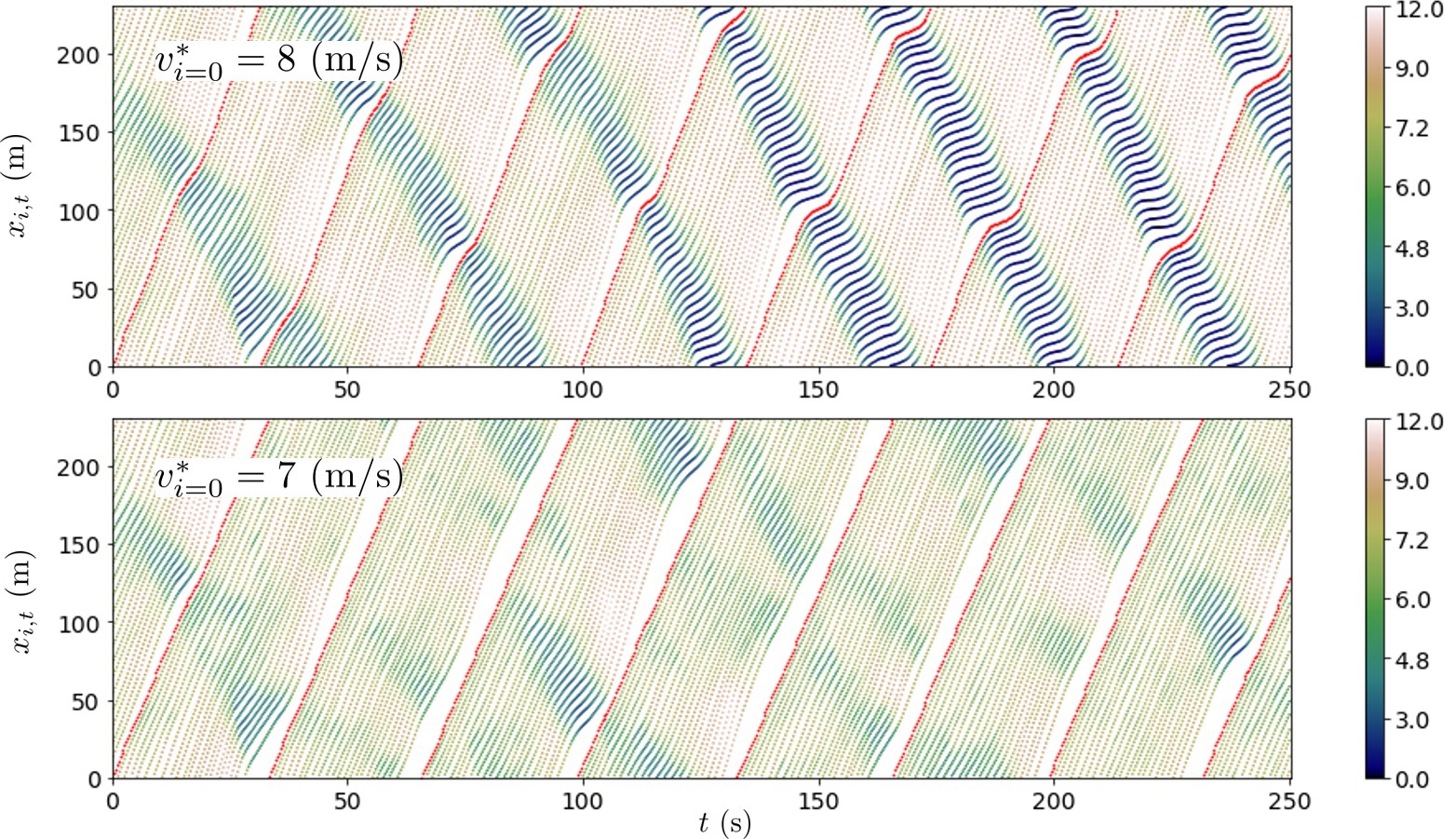}
\caption{Simulation results with vehicle $\#0$'s ideal speed lowered to 8 (m/s), the upper panel, and 7 (m/s), the lower panel. The red curve in either panel is associated with the trajectory of vehicle $\#$0.}
\label{Simulation_LowIdealSpeed}
\end{figure}

One possible mechanism for the shockwave formation is the chasing behaviors that each driver displayed during the course of Sugiyama experiment: the tendency to follow the vehicle ahead as close as safely possible. What happens if we break this tendency by forcing vehicle $\#0$ to have a sufficiently lower ideal speed than its nominal value? The simulation results are summarized in Fig.\ref{Simulation_LowIdealSpeed}. When $v_0^*=8$ (m/s) the shockwave can still be generated spontaneously, though not as easily as it used to be at higher ideal speed ($\sim 10$ m/s). As the ideal speed is further lower to $v_0^*=7$ (m/s), the shockwave becomes much harder to form spontaneously. Our simulation result is consistent with the work of \cite{Stern2018} in which a connected automated vehicle that aims at a constant speed adaptively was inserted into a string of human driven vehicles in a Sugiyama-like setting.
This forced non-chasing behavior suggests a possible mechanism to tame the shockwave systematically if we can control the ideal speed for a few CAV's in the midst of human driven vehicles at the system level. However, the truly relevant question here is not only to dissipate the traffic oscillation, but also to do so at reasonably high traffic flow. Otherwise, we could have forced all vehicles to stop and hence eliminated the oscillation. But this would also imply zero flow, something that is not desirable at all. Would it be possible to find Pareto improvements in such a way so that both traffic flow and smoothness become better relative to that in the jammed flow phase? How we make this idea concrete using optimal system control via computational mechanism design is investigated in \cite{Shen2020}, where we find that the answer to this question is gratifyingly affirmative.

\subsubsection{What if one vehicle has much higher risk premium}
\begin{figure}[!h]
\centering
\includegraphics[width=.80\textwidth]{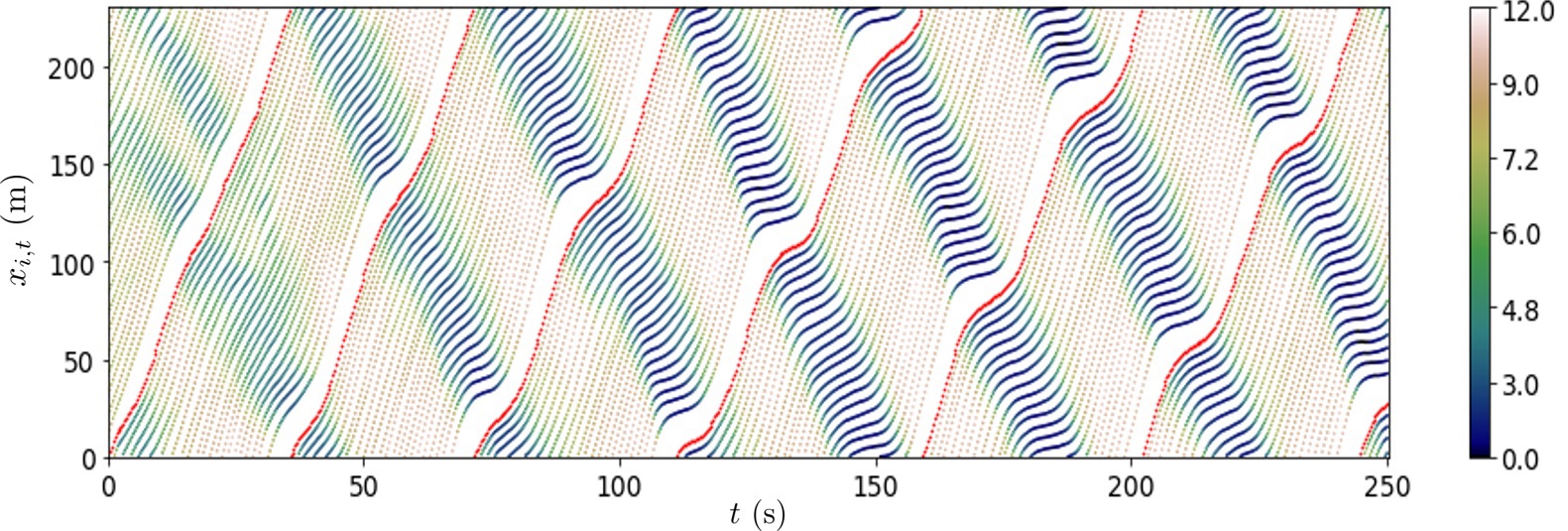}
\caption{Simulation results with vehicle $\#0$'s front risk premium: $\kappa_{0,v}^{(2,f)}=2.0$. The red curve is associated with the trajectory of vehicle $\#$0.}
\label{Simulation_HighRiskPremium}
\end{figure}

Would an extremely cautious driver also break the pattern of shockwave formation? To check out this possibility we force the risk premium parameter for vehicle $\#0$ to be very high: $\kappa_{0,v}=2.0$, i.e. many times larger than the average value. The simulation result is shown in Fig.\ref{Simulation_HighRiskPremium}. Interestingly, being extremely cautious is not enough to prevent shockwave being generated spontaneously. Apart from the fact that a big gap between vehicle $\#0$ and the vehicle ahead being obviously visible, all other characteristics are more or less intact. This result further strengthens the belief that the spontaneous formation of shockwave requires chasing behavior for all local participants.

\subsection{Simulations in Tadaki Setting}\label{SimulatedTadaki}
We also simulated the Tadaki experiment setting, which uses a circular road with 314 m circumference and varying number of identical cars \cite{Tadaki13}. We were able to re-produce the non-trivial vehicle density dependence for the fundamental diagram and the critical region of vehicle density, which separates the free flow phase from the jammed flow phase. The detailed results will be presented in \cite{Shen2020}, along with addressing the issue of how to tame stop-and-go shockwaves using system controlled connected vehicles in an optimal way via computational mechanism design.

\subsection{Simulated single-lane highway}\label{SimulatedHighway}
\begin{figure}[!h]
\centering
\includegraphics[width=.80\textwidth]{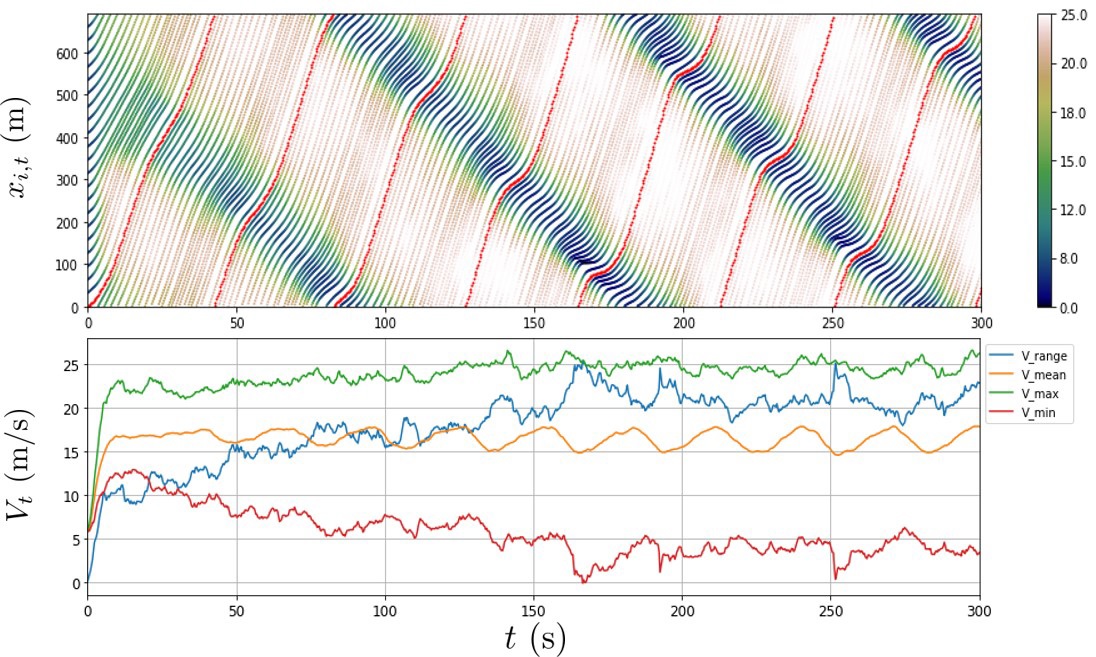}
\caption{Simulated traffic in a simple one-lane highway setting. The red curve in the upper panel is associated with the trajectory of vehicle $\#$0.}
\label{SimulatedOneLaneHighway}
\end{figure}

Before we finish the work for calibrating a behavioral model on highway systematically using NGSIM or HighD data in \cite{Shen2021}, could we glean anything useful at all with what we already got so far? In this subsection we scale up the circumference of the Sugiyama experiment by a factor of 3, with the number of vehicles kept fixed ($N=22$). This scaling obviously imply that we should also scale up the ideal speed by the same factor, which in turn would allow us to simulate the traffic at an average speed at about three times as fast as in the original Sugiyama setting, similar to a single-lane highway setting. Less obvious is how we should scale the other model parameters in this drastically new setting. From Fig.1 in \cite{Tadaki13} and Fig.9 in \cite{Nakayama2016}, we know that the shockwaves start roughly at the traffic density of $25$ cars/km, or at an average headway of about 40 m, with average speed of roughly 25 to 30 m/s. This gives us a clue that the headway related risk premium parameters, i.e. $\kappa_v$, should be scaled up by roughly factor of 3 also.\footnote{This is also consistent with the fact that the distance required for a complete stop of a vehicle with constant braking is quadratically proportional to its initial speed.} Likewise, $\kappa_c$ perhaps should be scaled by a factor of 3 to 9 (we chose 5). Would shockwaves be spontaneously generated in our model with the scaled parameters, while leaving all other model parameters unchanged? We show the simulation result in Fig.\ref{SimulatedOneLaneHighway}, where shockwaves are clearly visible, broadly consistent with what's observed in Fig.1 of \cite{Tadaki13} and and Fig.9 in \cite{Nakayama2016}. Even more interesting is the similarity of our simulation results to NGSIM data (see the congested traffic video of US101 from 8:05 AM to 8:20 AM): 1) the backward traveling speed of the shockwave is around 5 (m/s); and 2) the queue length is about 10 cars per slow-moving cluster.

This semi-quantitative consistency, along with the encouraging quantitative simulation results in Tadaki setting described in section \ref{SimulatedTadaki}, implies that our model could be generalized reasonably well to new settings in an intuitive manner, providing strong evidences for the common belief that it is much easier to generalize using an objective-based approach than a policy-based approach. Of course, this is only two simple examples, and more systematic and careful studies are definitely needed for firmer conclusion on generalizability. 

\section{\bf Summary and Conclusions}\label{SummaryConclusions}
In this paper we have developed a systematic framework for modeling human driving behaviors using naturalistic traffic data. Our method leverages the recently proposed heuristics based decision-making model that was simplified from a Markov game framework \cite{Dai2020}. We explicitly demonstrated that the forward decision-making model can be inverted using individual vehicle trajectory data from Sugiyama experiment. In our opinion, all the design goals listed in Introduction were met reasonably well. Our simulation results based on the estimated model also confirmed qualitatively many of the findings in a lot of car-following models, especially in a recent work \cite{Treiber2017}.

There are a number of directions we can push the methodology further. Attentions can be shifted to more sophisticated urban settings that are still relatively simple, such as single-lane roundabout and four-way intersection, where naturalistic vehicle trajectories and relevant surrounding info can be captured by using drone. Interesting new phenomena that were missing in the Sugiyama case, such as merging and interacting with traffic signals, will become central. One slight complication arising from these cases is that the state evolution has to be based on the kinematic bicycle model \cite{Rajamani06}, which will render the state space model nonlinear that needs to be handled by an extended Kalman filter. Other obvious places where we can apply our methodology are highways with or without ramps. 

Two more methodological issues need to be taken care of explicitly for these new applications. First, the observation of a single vehicle in a drone video is typically too short, at an order of around 30 to 60 seconds, to provide sufficient sampling of the necessary regions of the state space required by the inversion algorithm. This may imply that inference at individual level is not possible with videos captured by a single drone. We either have to fuse data from multiple drones or to cleverly combine data from multiple vehicles with similar yet unknown driving characteristics so that heterogeneity still can be modeled at group level. Second, unrealized driving intentions may not be observable from drone video data. One such example is when a vehicle intended to exit from a ramp but was blocked due to heavily congested traffic. One possible way to solve both of these issues is to resort to the technique of latent class modeling as pointed out in \cite{Babes2011}. We are actively pursuing this line of inquiry in \cite{Song2021, Shen2021}. 

Another potentially fruitful exercise is to extend the inference algorithm presented here to online environments so that real-time learning of individual driving style adapted to the specific driving scenario could become possible. Of course, the data requirement will be extremely high, necessitating very long trajectories for each vehicle/driver along with its state information.

\section*{Acknowledgement} We would like to thank Prof. Sugiyama and his collaborators for sharing their experiment data with us. We also express our gratitude to Alan Xu and Wen Guo for their initial participation of this work.

\bibliographystyle{unsrtnat}
\bibliography{calibration_refs}

\end{document}